\documentclass[twocolumn]{revtex4}
\usepackage{amssymb}
\usepackage{latexsym}
\usepackage{graphicx}
\usepackage{epsfig}

\usepackage{amsmath}
\usepackage{latexsym,graphicx,multirow}
\usepackage{amssymb}
\usepackage{amsmath}
\usepackage{amscd}
\usepackage{amsthm}

\begin{document}
	\title{ \bf Comparing holographic principle inspired dark energy models}
	\author{Vipin Chandra Dubey$^{1}$\footnote{vipin.dubey@gla.ac.in},  Umesh Kumar Sharma$^{2}$\footnote{ sharma.umesh@gla.ac.in}}
	\address{$^{1,2}$ Department of Mathematics, Institute of Applied Sciences and Humanities, GLA University, Mathura, U. P., 281406, India\\}

	\begin{abstract}
	In this article, we have analysed holographic principle inspired dark energy models, namely Holographic dark energy (HDE) model, Tsallis Holographic dark energy (THDE) model and the R$\acute{e}$nyi holographic dark energy (RHDE) model by using the statefinder parameter diagnostic. We have compared these dark energy models by plotting the evolutionary trajectories of the first statefinder $r(z)$, second statefinder parameter $s(z)$, the statefinder parameter pairs $(r, s)$ and $(r, q)$ as well as the deceleration parameter $q(z)$ for the various parameter values of the respective dark energy models. It is observed from $q(z)$, $r(z)$ and $s(z)$ plots that the THDE model and RHDE model are not discriminated from $\Lambda$CDM model at the low redshift region, while the HDE model is well-differentiated from $\Lambda$CDM model at the low redshift region. Clear discrimination among these three dark energy models as well as from the $\Lambda$CDM model can be seen in the $(r, s)$ and $(r, q)$ planes.
	\end{abstract}

	\maketitle

	\section{Introduction}

Presently, the expansion of our universe is accelerating, which is verified by various cosmological observations \cite{ref1,ref2,ref7,ref8,ref9,ref10}.The dark energy (DE) is introduced for explaining this accelerated expansion in the framework of the standard cosmology, which is an obscure component with negative pressure \cite{ref10a,ref10b,ref12,ref13,ref13a}. Among various kinds of DE models, the cosmological constant model i.e. $\Lambda$CDM model is the simplest one \cite{ref14a,ref14b,ref14c,ref14d,ref14e}. For $\Lambda$CDM model, time - independent equation of state (EoS) $\omega_{\Lambda} = - 1$, but it faces theoretical problems, specifically the fine tuning problem as well as the coincidence problem \cite{ref14b,ref14f,ref14g}.  To get alleviation from such issues, numerous dynamical DE models are given as a choice such as  tachyon \cite{ref14m}, quintessence \cite{ref14i,ref14j}, $k$-essence \cite{ref14h,ref14h1}, phantom \cite{ref14l} and Chaplygin gas \cite{ref14k}.\\	
	
	Inspired by the holographic principle (HP) \cite{ref22,ref23,ref24,ref25}, the holographic dark energy (HDE) model was proposed by Li \cite{ref26} in 2004. According to  HP, the number of degrees of freedom for a system does not depend on the volume but its bounding area. The accelerated expansion phase of the Universe can be explained by the HDE \cite{ref27,ref28}. In anticipation of quantum modification for HDE, Tsallis and Cirto produce Tsallis entropy, which is a generalization of the Boltzmann-Gibbs (BG) entropy to non-extensive systems\cite{ref29,ref29a1,ref29a2}.  Being peculiar behaviour of the Universe and long-range aspects of gravity. A generalized entropy formalism has been considered for investigating the peculiar behaviour of the Universe and long-range aspects of gravity, which inspires us to a decent consent with gravity and its associated problems \cite{ref30,ref31}.\\
	
	Different entropies are also used for the investigation of the gravitational and cosmological scenario. Recently, two new forms of DE models based on HP, Tsallis and R$\acute{e}$nyi \cite{ref31a,ref31b} entropy are proposed by Tavayef et. al \cite {ref32} and by Moradpour et. al \cite {ref33}, known as  Tsallis holographic dark energy (THDE) and R$\acute{e}$nyi holographic dark energy (RHDE), respectively. Both the THDE and RHDE might be consistently called HP inspired dark models for simplicity. These models of  HDE can be used to clarify or explain the cosmic acceleration of the universe \cite{ref34,ref35,ref36,ref37,ref38,ref39,ref40,ref41,ref42,ref43,ref44,ref45,ref46,ref47,ref48,ref49,ref50,ref51}. \\
	
The DE phenomenon can be explained by many DE models.  Hence it becomes very important to distinguish among various vying DE models. In this way, Sahni et al. \cite{ref52} and Alam et al. \cite{ref53} introduced a significant geometrical indicative, namely statefinder pair (${r, s}$) for removing the degeneracy in present values of $q$ and $H$ for different DE models. It is also used widely to recognise different models of modified theories of gravity.  Since we get distinct evolutionary trajectories in $(s, r)$ plane for different DE models. The statefinder parameters are analysed in \cite{ref54,ref55,ref56,ref57,ref58,ref59,ref60,ref61}. The statefinder parameter diagnostic is also used to discriminate various DE models such as the Ricci dark energy (RDE), the new HDE,  the new ADE and the original HDE model \cite{ref63}.\\
	
	Motivated by the work of ref. \cite{ref63}, in this paper, we compare the original
HDE model  with some newly proposed DE models such as the Tsallis holographic
	dark energy (THDE) model and R$\acute{e}$nyi holographic dark energy (RHDE) model through the statefinder parameters $(r-s)$ diagnostic.
	In
	Sect. 2,  we reviewed briefly  the HP inspired DE  models.
	In
	Sect. 3 we  diagnosed these HDE models with the deceleration parameter $q$,
	and the statefinder. The
	conclusion is given in Sect. 4.

	\begin{figure}[htp]
		\begin{center}
			\includegraphics[width=9cm,height=8cm]{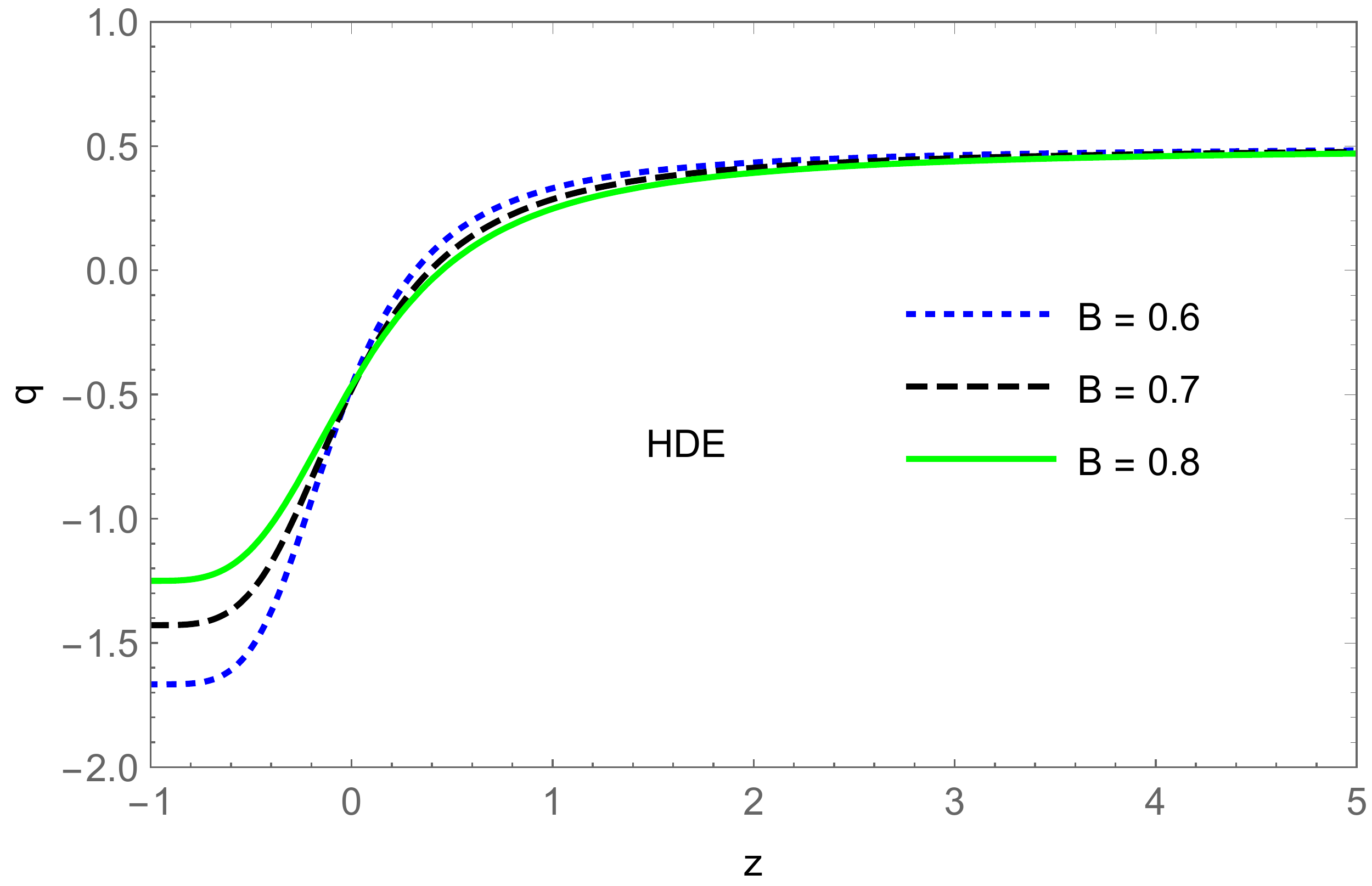}
			\includegraphics[width=9cm,height=8cm]{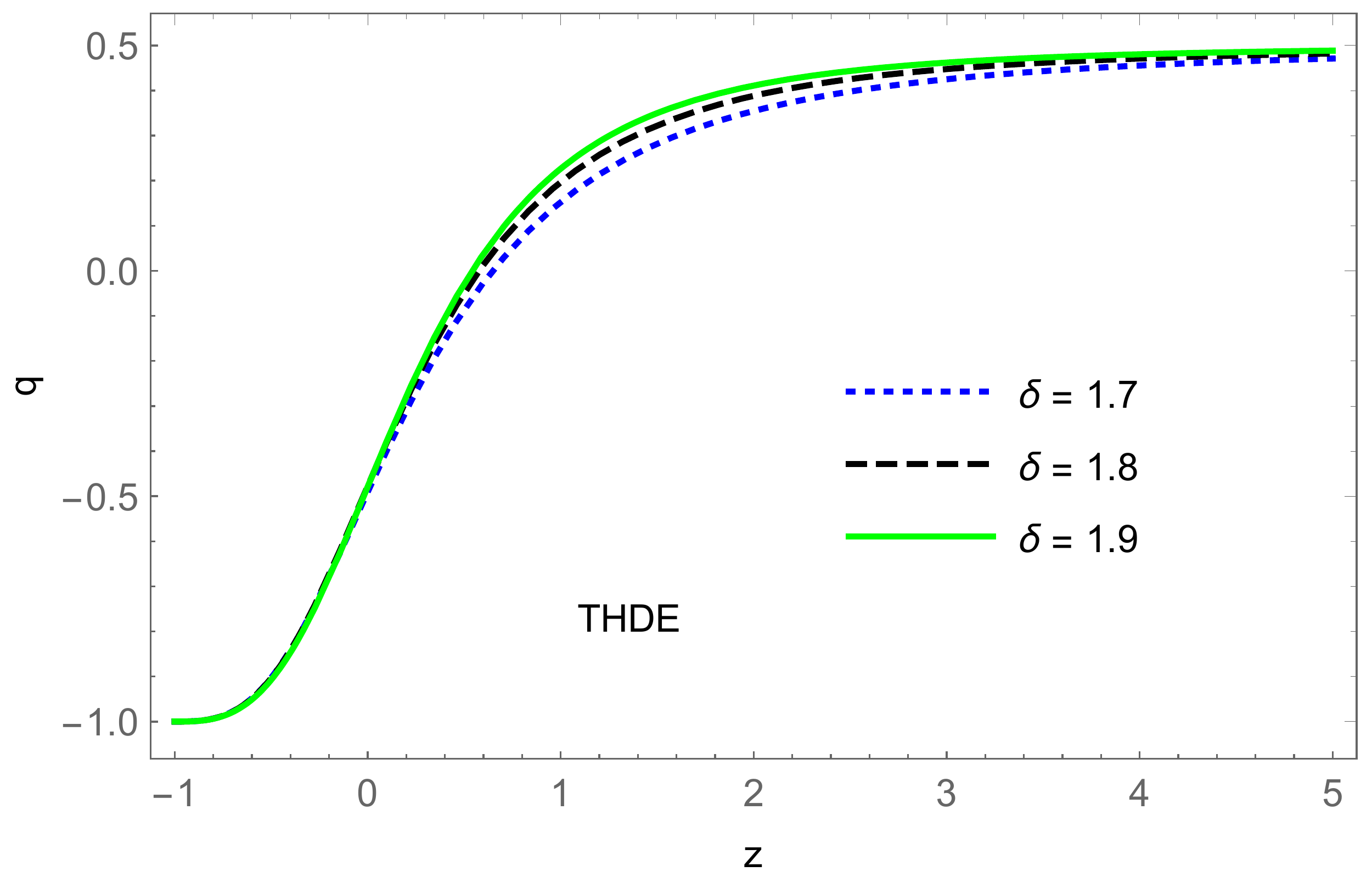}
			\includegraphics[width=9cm,height=8cm]{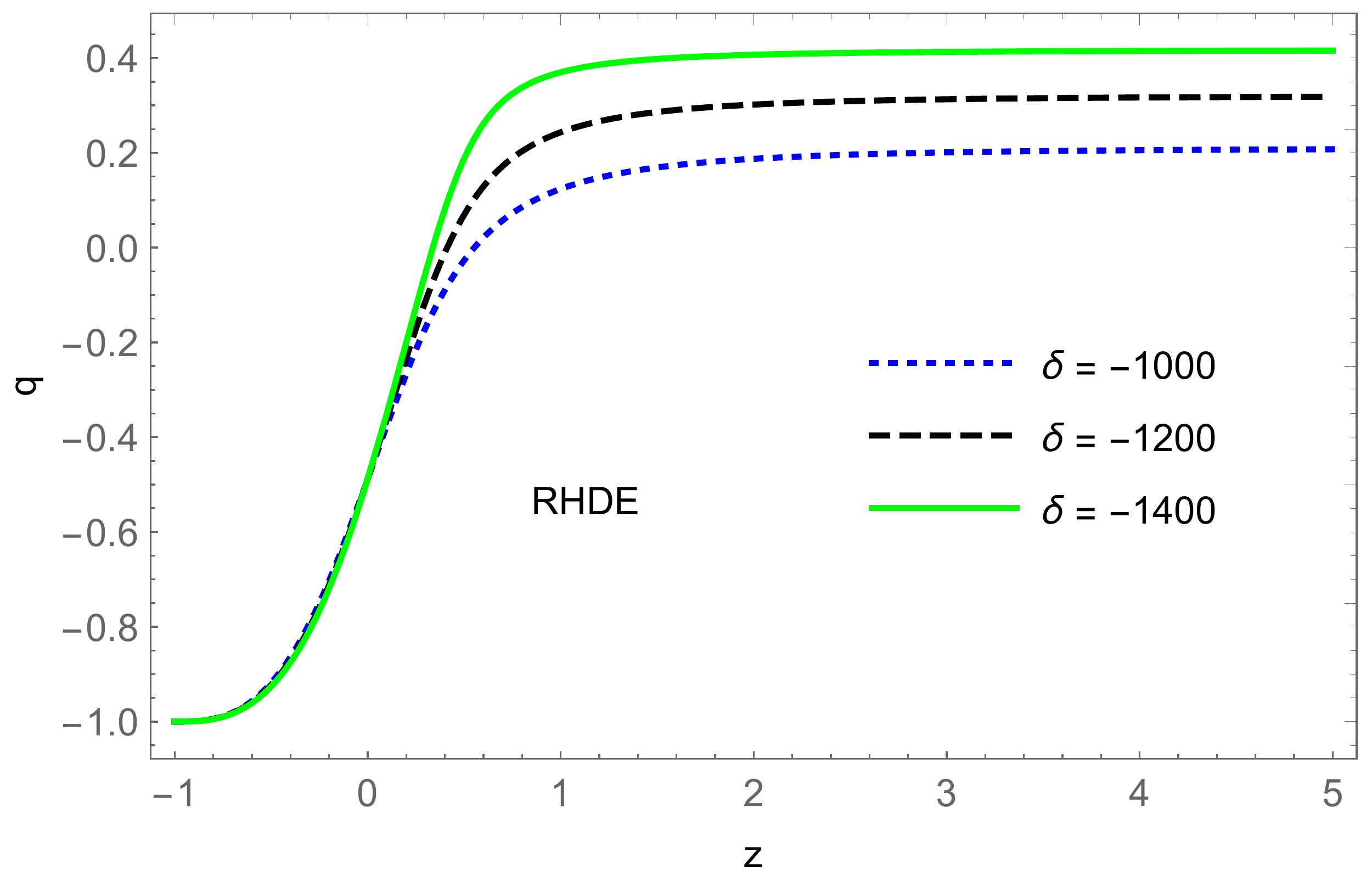}
			\caption{The evolutionary behaviours of the  $q$ (deceleration parameter)  versus  $z$ (redshift)  for HDE, THDE and RHDE Models. Here, $\Omega_D(z=0)=0.73$, $H(z=0)= 70$, $\omega_{D}(z=0) = -
				0.90$}
			\label{figure 1}
		\end{center}
	\end{figure}
	
	\begin{figure}[htp]
		\begin{center}
			\includegraphics[width=9cm,height=8cm]{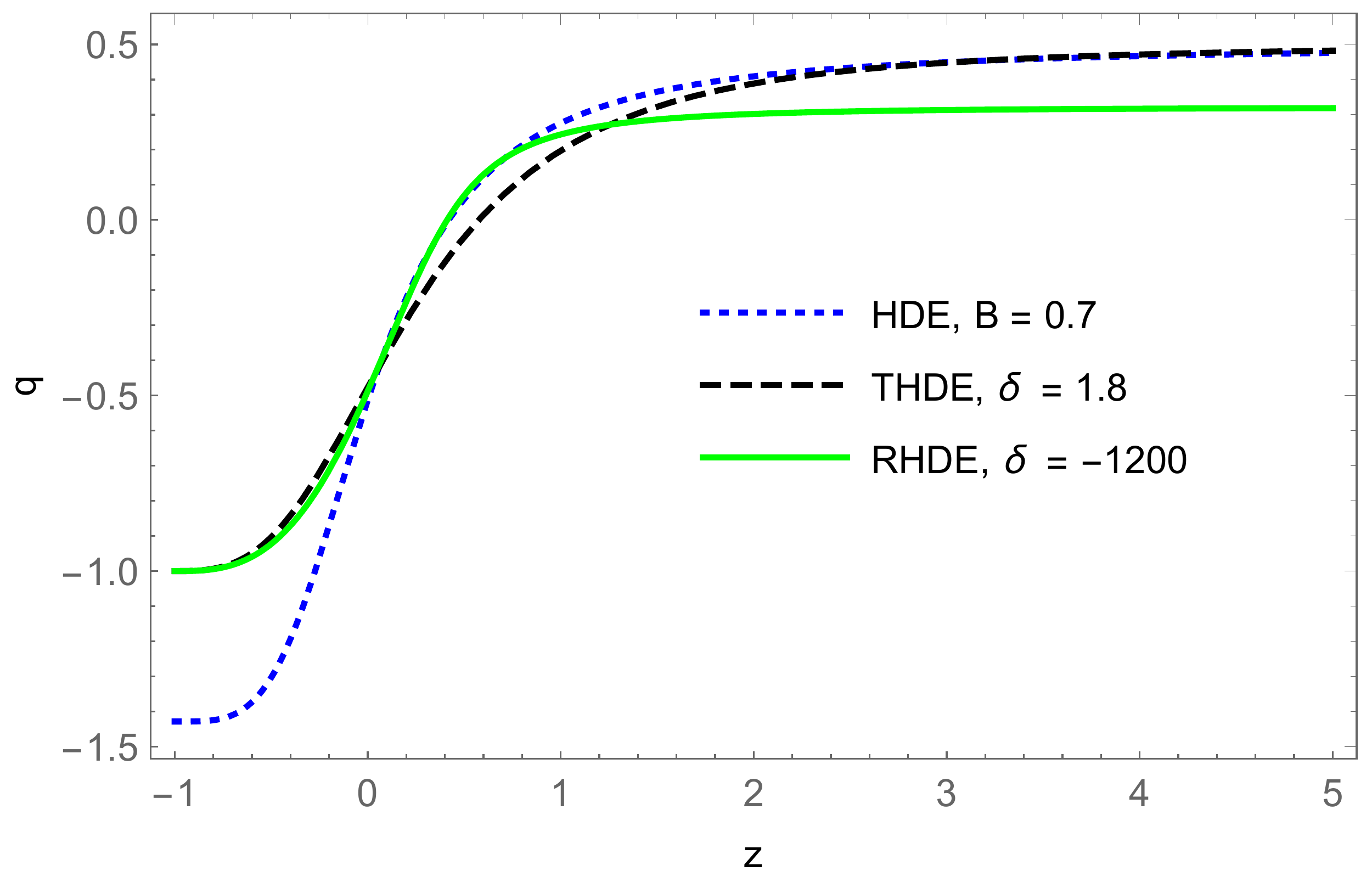}
			\caption{ Comparison of evolutionary behaviour of $q$ for the HP inspired DE models.   Here, $\Omega_D(z=0)=0.73$, $H(z=0)= 70$, $\omega_{D}(z=0) = -
				0.90$}
			\label{figure 2}
		\end{center}
	\end{figure}

	\section{Holographic principle inspired DE models}
	
	\subsection{The HDE model}
	Considering a spatially flat Friedmann Robertson Walker Universe accommodating matter and dark
	energy (assuming a flat Universe in the
	complete manuscript), the Friedmann equation is given as
	
	\begin{equation}
	\label{eq1}
	H^{2} = \frac{1}{ 3 M^{2}_{p} } \left( \rho_{m} + \rho_{D}\right),
	\end{equation}
	
	where $\rho_{m} $,  $\rho_{D} $ represent the  energy densities for matter and  DE, respectively, and   $M_{p}$= $\frac{1}{\sqrt{8 \pi G}}$ is the reduced Planck mass. $H$ is the Hubble parameter.  For the HDE model \cite{ref22,ref63a}, $ \rho_{D} = 3B^{2}M^{2}_{p}L^{-2}$, where $L$ is the largest IR cut-off
		 and $B$ is a dimensionless constant.  Now,

		 by taking the future event horizon as IR cut-off $L$, we have:
	
	\begin{equation}
	\label{eq2}
	R_{h}=a\int _a^{\infty }\frac{{da}'}{H(a') a'^2},
	\end{equation}
	
 Eq. (\ref{eq1}) can be written as: 
		\begin{equation}
	\label{eq2a}
1=\Omega _D + \Omega _m,
	\end{equation}
	Where  $\Omega _m =  \Omega^{0} _m H^{2} _0 H^{-2} a^{-3}$
	and  $\Omega _ D=  B^{2} H^{-2}R_{h}^{-2} $
	 be the relative energy densities of matter and DE, respectively, which are expressed as fractions of the critical density  $ \rho_{c} = 3M^{2}_{p}H^{2}$. Here, $H _0$ and $\Omega _{{m0}}$ are the present values of the Hubble parameter $H$ and the  matter density parameter $\Omega _{{m}}$. Now, by using $R_{h} = B/H \sqrt{\Omega _ D}$ and  Eq. (\ref{eq2}), we have:\\
 	\begin{equation}
 \label{eq2b}
\int _x^{\infty }\frac{{da}'}{H(a') a'}= \frac{B}{H(a) a \sqrt{\Omega _ D} },
 \end{equation} \\
where $x = log a $. Now from the Friedmann equation, we have \cite{ref56} :\\
 \begin{equation}
 \label{eq2c}
 \frac{1}{H(a) a}=\sqrt{a (1 -\Omega _ D)}  \frac{1}{H _0  \sqrt{\Omega^{0} _ m}},
 \end{equation} \\ 
From Eq. (\ref{eq2b}) and Eq. (\ref{eq2c}), we get the relation
	\begin{equation}
\label{eq2d}
\int _x^{\infty } e^{x'/2} \sqrt{1 - \Omega _ D} dx' = B e^{x/2}  \sqrt{\frac{1}{\Omega _ D} -1}.
\end{equation} \\
Now, differentiating Eq. (\ref{eq2d}) with respect to $x = log a $, we have	
		\begin{equation}
	\label{eq3}
	\Omega _D'=\Omega _D \left(1-\Omega _D\right) \left(\frac{2 \sqrt{\Omega _D}}{B}+1\right).
	\end{equation}
	
	Where the prime represents  the differential coefficient  with 
	$x = log a $. Also, the energy conservation law, $\dot \rho_{D} + 3H (\omega _D  + 1) \rho_{D} = 0$, the equation of state (EOS) parameter of
the	HDE, $ \omega _D\equiv p _D/\rho _D$, can be given by \cite{ref56}:
	
	\begin{equation}
	\label{eq4}
	\omega _D= -\frac{1}{3} -\frac{2 \sqrt{\Omega _D}}{3B},
	\end{equation}
	\begin{figure}[htp]
		\begin{center}
			\includegraphics[width=9cm,height=8cm]{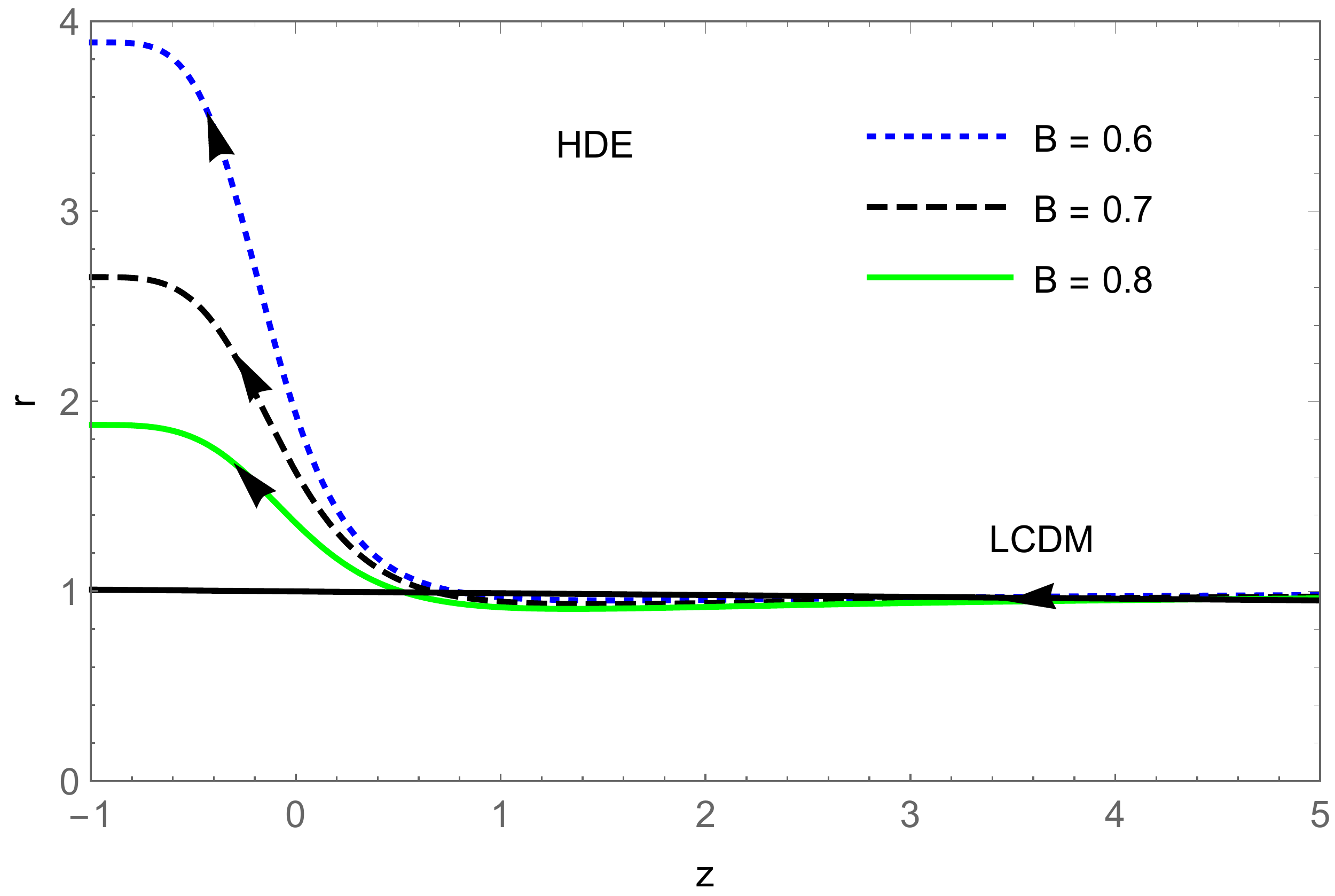}
			\includegraphics[width=9cm,height=8cm]{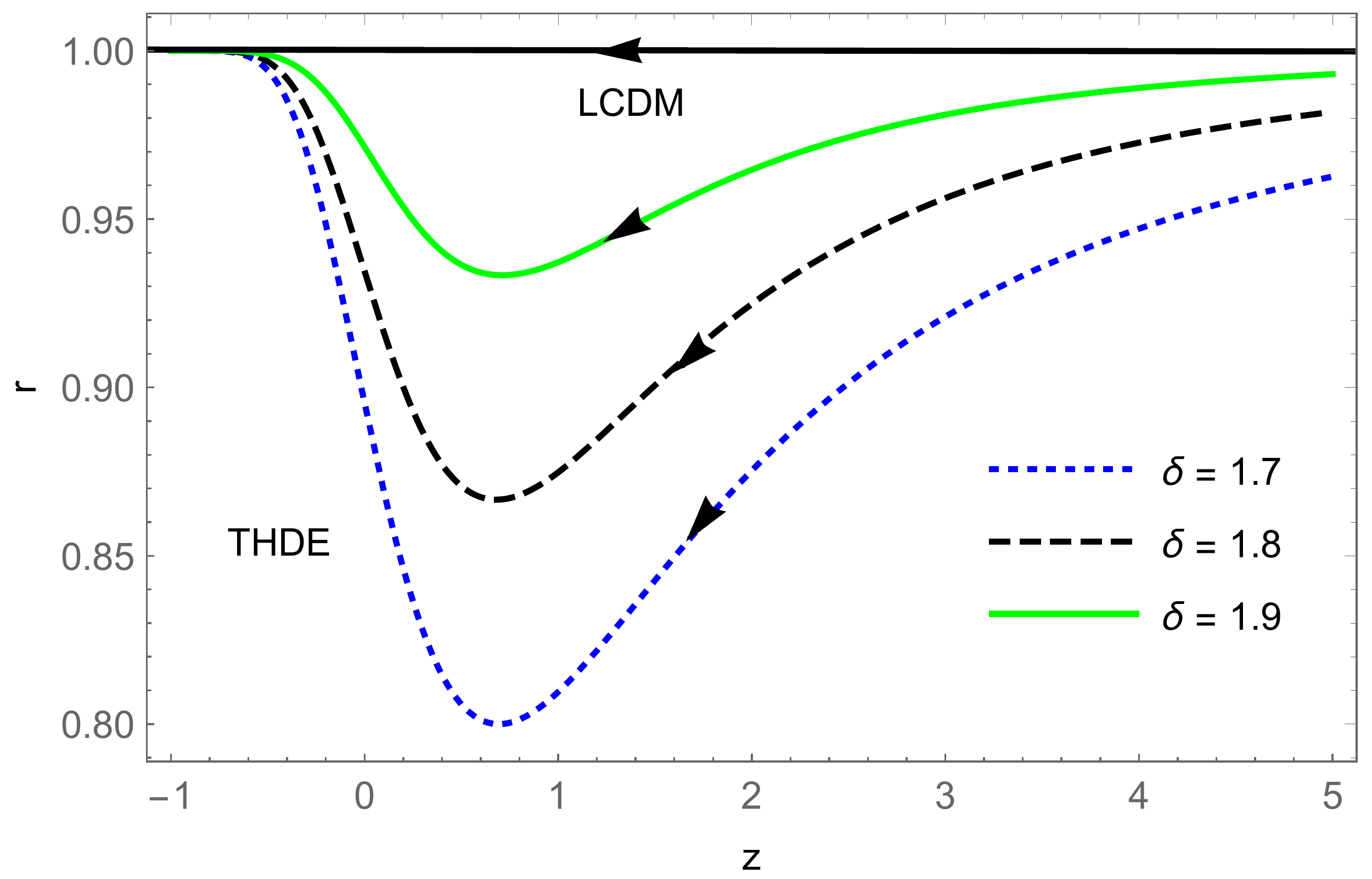}
			\includegraphics[width=9cm,height=8cm]{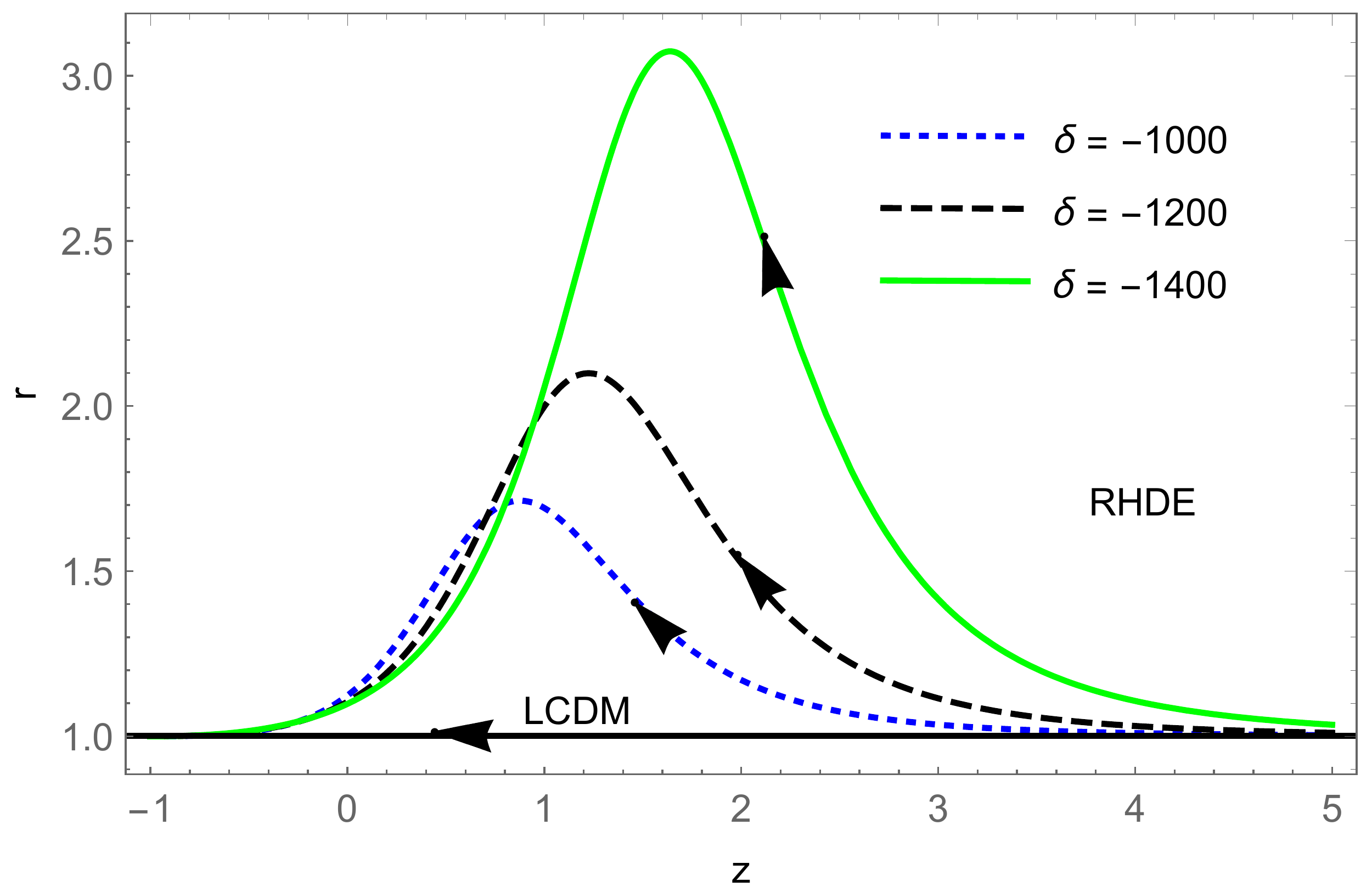}
			\caption{The evaluation of the first statefinder parameter $r$ versus redshift parameter $z$ for HDE, THDE and RHDE Models.  Here, $\Omega_D(z=0)=0.73$, $H(z=0)= 70$, $\omega_{D}(z=0) = -
				0.90$.}
			\label{Omega-z1}
		\end{center}
	\end{figure}
	
	\begin{figure}[htp]
		\begin{center}
			\includegraphics[width=9cm,height=8cm]{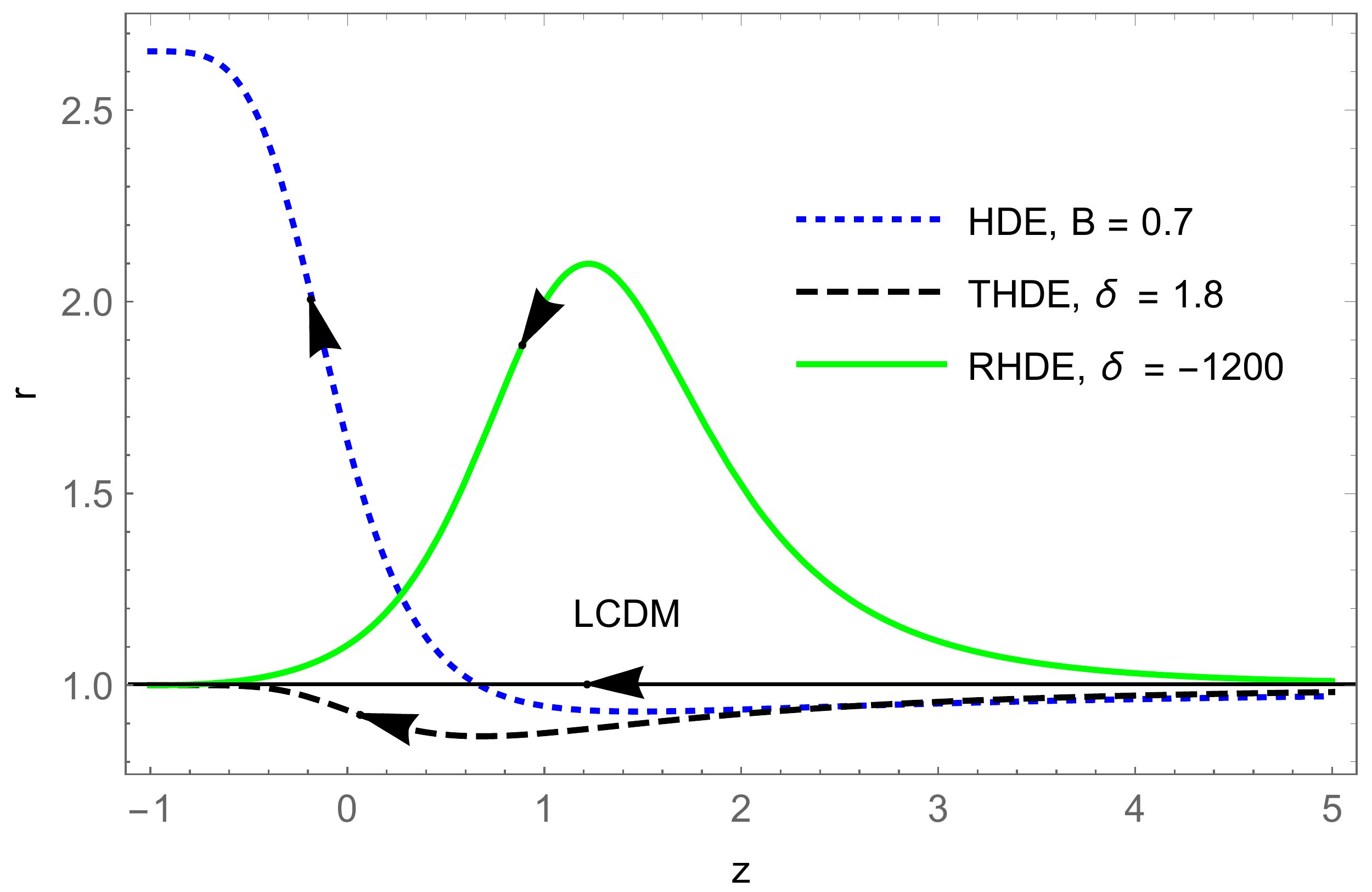}
			\caption{  Comparison of evolutionary trajectories  of the first statefinder parameter $r$ for the HP inspired DE models.   Here, $\Omega_D(z=0)=0.73$, $H(z=0)= 70$, $\omega_{D}(z=0) = -
				0.90$.}
			\label{Omega-z2}
		\end{center}
	\end{figure}
	
	\begin{figure}[htp]
		\begin{center}
			\includegraphics[width=9cm,height=8cm]{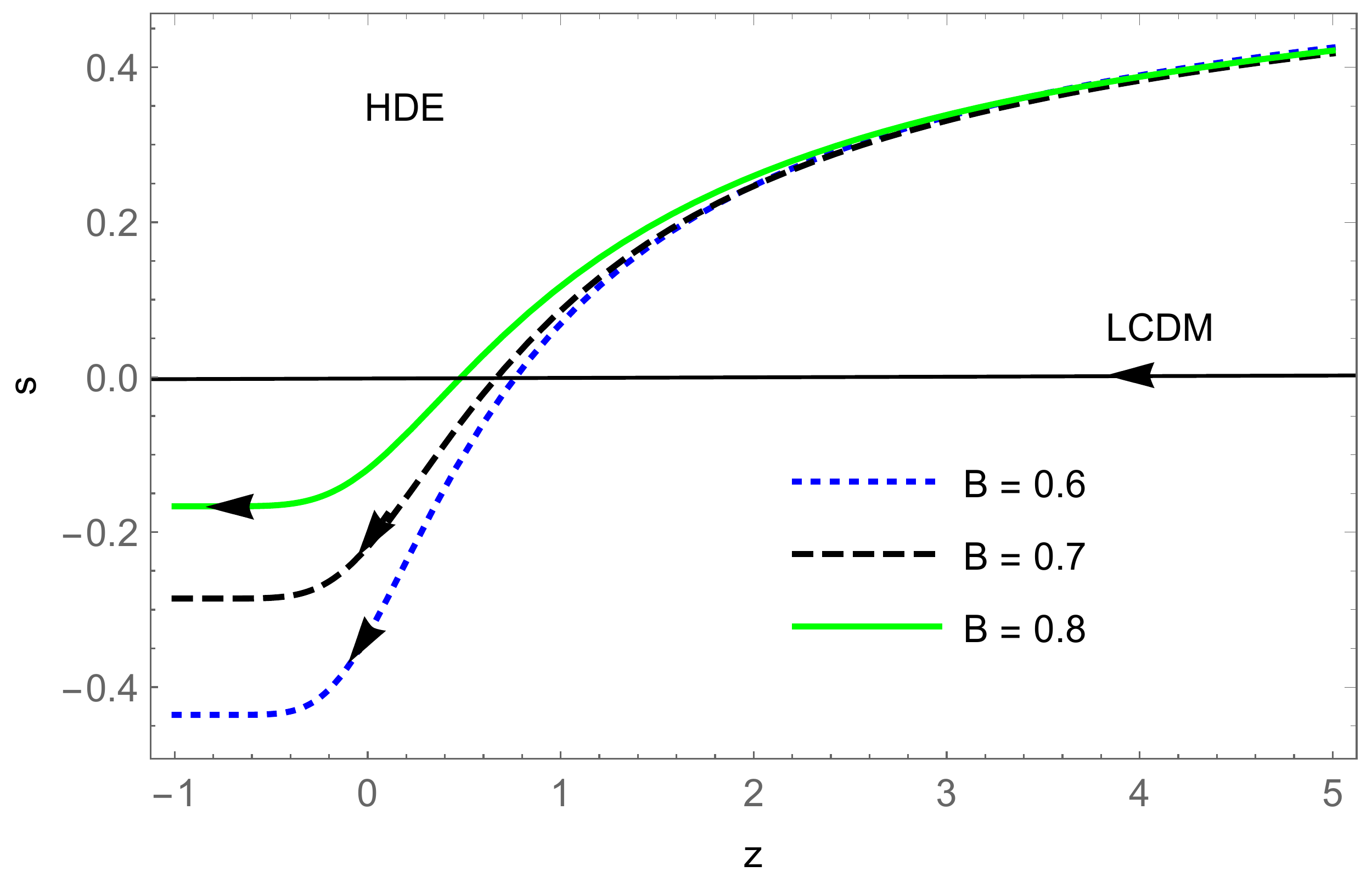}
			\includegraphics[width=9cm,height=8cm]{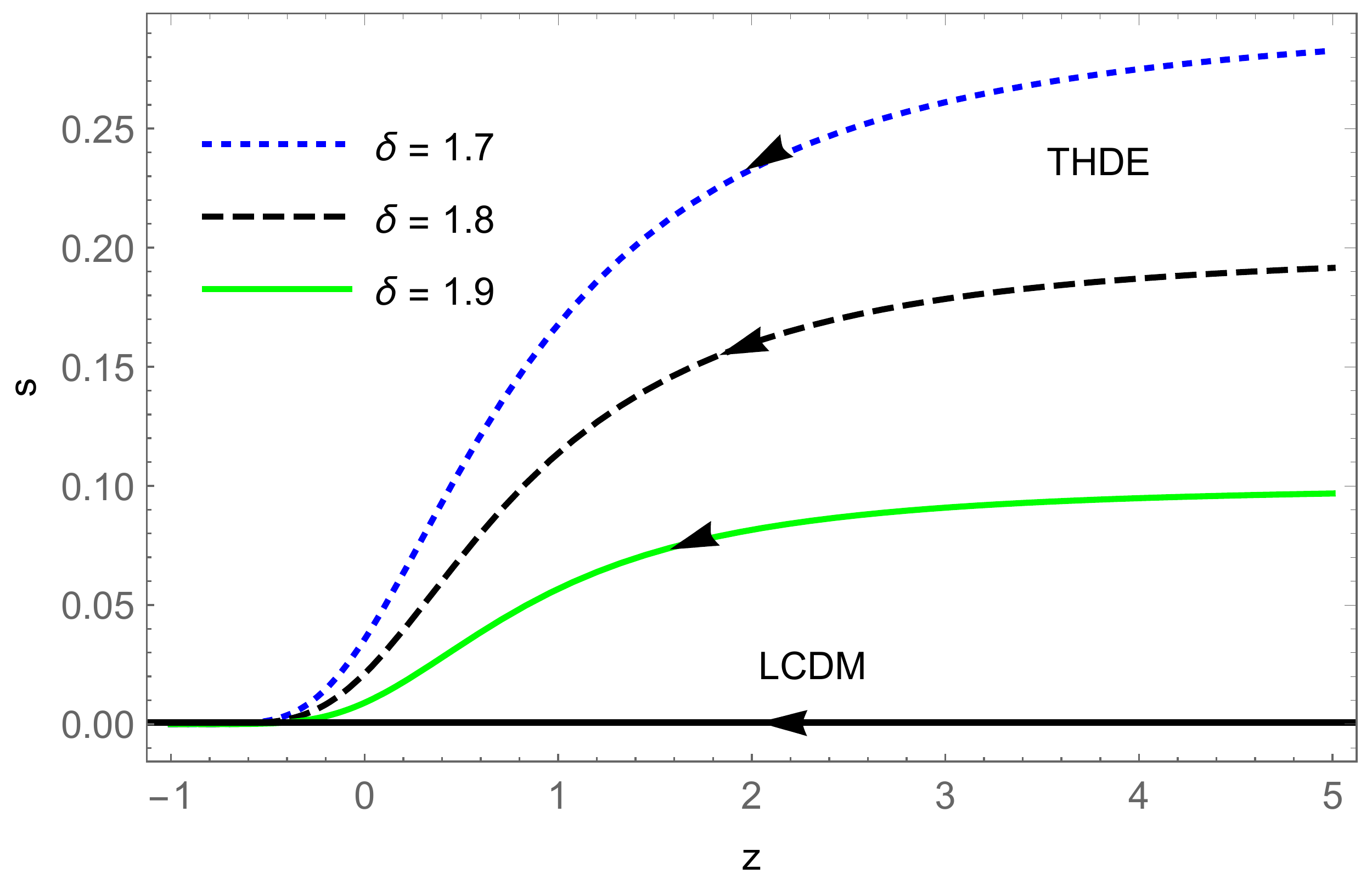}
			\includegraphics[width=9cm,height=8cm]{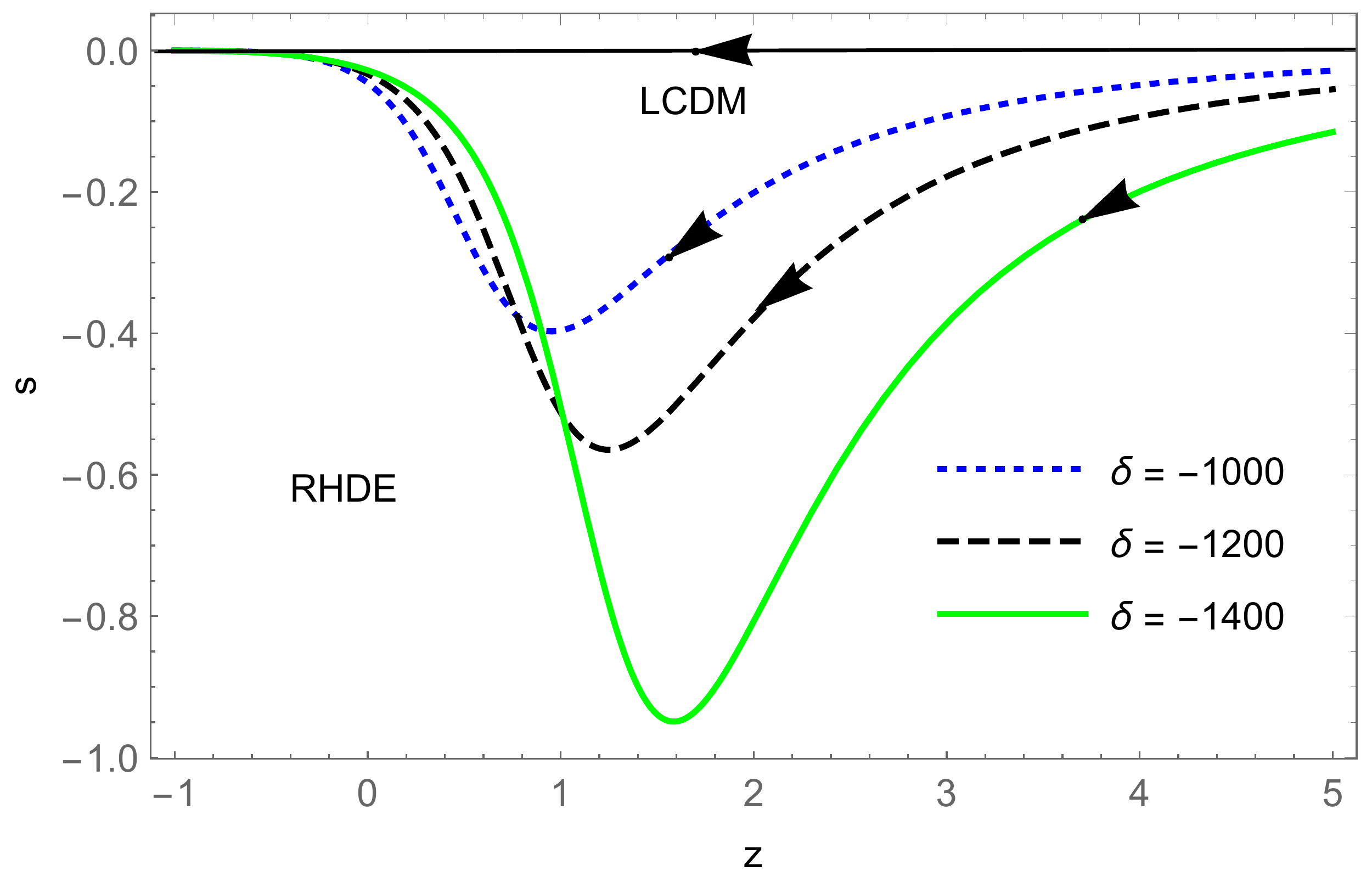}
			\caption{The evolutionary trajectories of the second statefinder parameter $s$ versus redshift parameter $z$ for HDE, THDE and RHDE models.  Here, $\Omega_D(z=0)=0.73$, $H(z=0)= 70$, $\omega_{D}(z=0) = -
				0.90$.}
			\label{Omega-z3}
		\end{center}
	\end{figure}
	
	\begin{figure}[htp]
		\begin{center}
			\includegraphics[width=9cm,height=8cm]{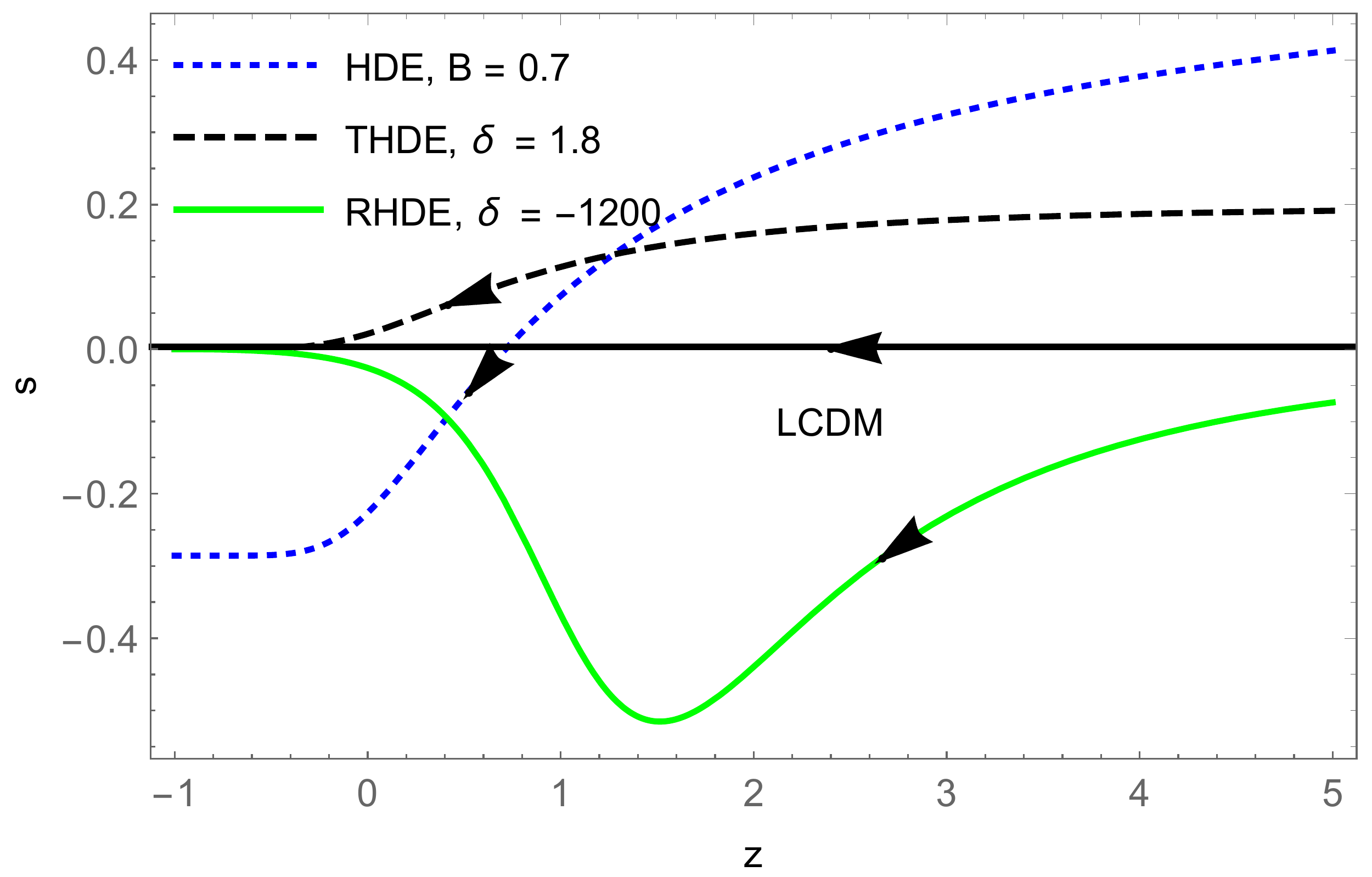}
			\caption{ Comparison of evolutionary trajectories  of the second statefinder parameter $s$ for the HP inspired DE models.   Here, $\Omega_D(z=0)=0.73$, $H(z=0)= 70$, $\omega_{D}(z=0) = -
				0.90$.}
			\label{Omega-z4}
		\end{center}
	\end{figure}
	
	\subsection{The THDE model} 
	Newly, another form of non-extensive and HP inspired    dark model  has been  proposed  in \cite{ref32}, and the  energy density for the THDE is defined as
	
	\begin{equation}
	\label{eq5}
	\rho _D=B H^{4-2 \delta },
	\end{equation}
	
	where the parameter $B$ is unknown and $\delta$ is the non-extensive real parameter, which quantifies the degree of non-extensive \cite{ref29,ref29a1,ref30,ref33e}. 	Taking the time derivative of $\Omega_D \equiv \frac{\rho_{D}}{ 3 M^{2}_{p}H^{2}}$ on both sides with respect to $x = log a $, we get
	
	\begin{equation}
	\label{eq6}
	\Omega _D'=\frac{3 (\delta -1) \Omega _D \left(1-\Omega _D\right)}{1-(2-\delta ) \Omega _D},
	\end{equation}
	So, the THDE equation of state (EoS) parameter  is inferred as
	
	\begin{equation}
	\label{eq7}
	\omega _D=\frac{\delta -1}{(2-\delta ) \Omega _D-1}
	\end{equation}
	
	\begin{figure}[htp]
		\begin{center}
			\includegraphics[width=9cm,height=8cm]{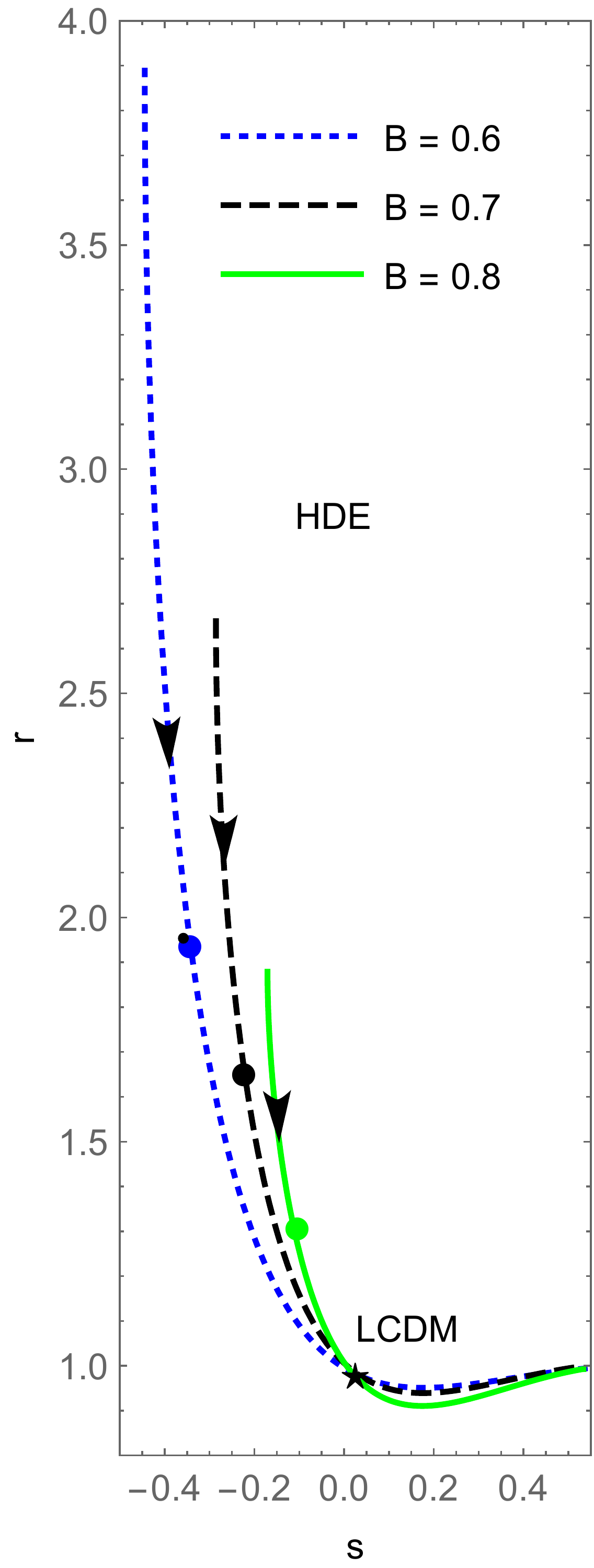}
			\includegraphics[width=9cm,height=8cm]{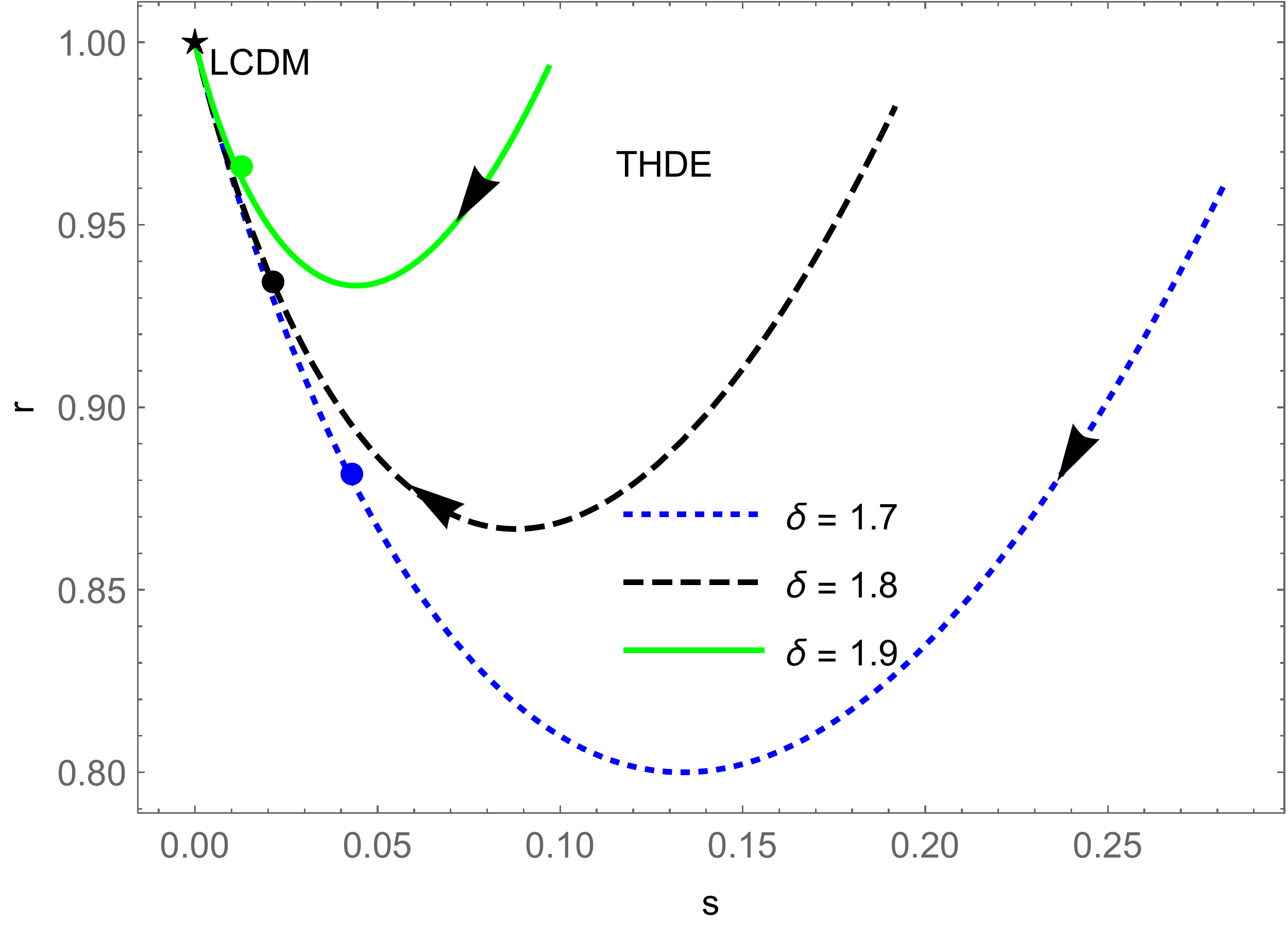}
			\includegraphics[width=9cm,height=8cm]{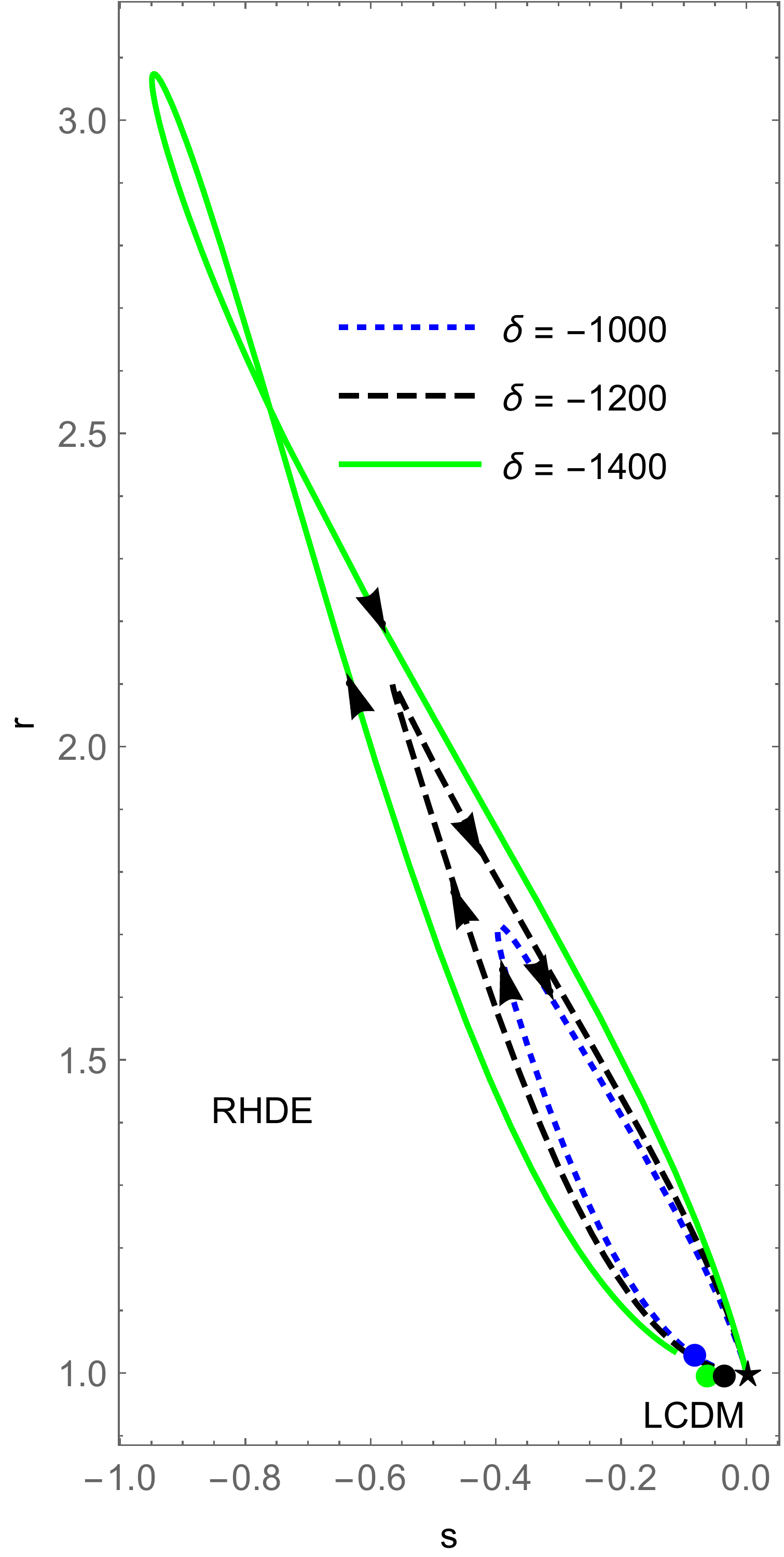}
			\caption{The evolutionary trajectories of $r$ versus $s$ for HDE, THDE and RHDE models.  Here, $\Omega_D(z=0)=0.73$, $H(z=0)= 70$, $\omega_{D}(z=0) = -
				0.90$.}
			\label{Omega-z7}
		\end{center}
	\end{figure}
	
	\begin{figure}[htp]
		\begin{center}
			\includegraphics[width=9cm,height=8cm]{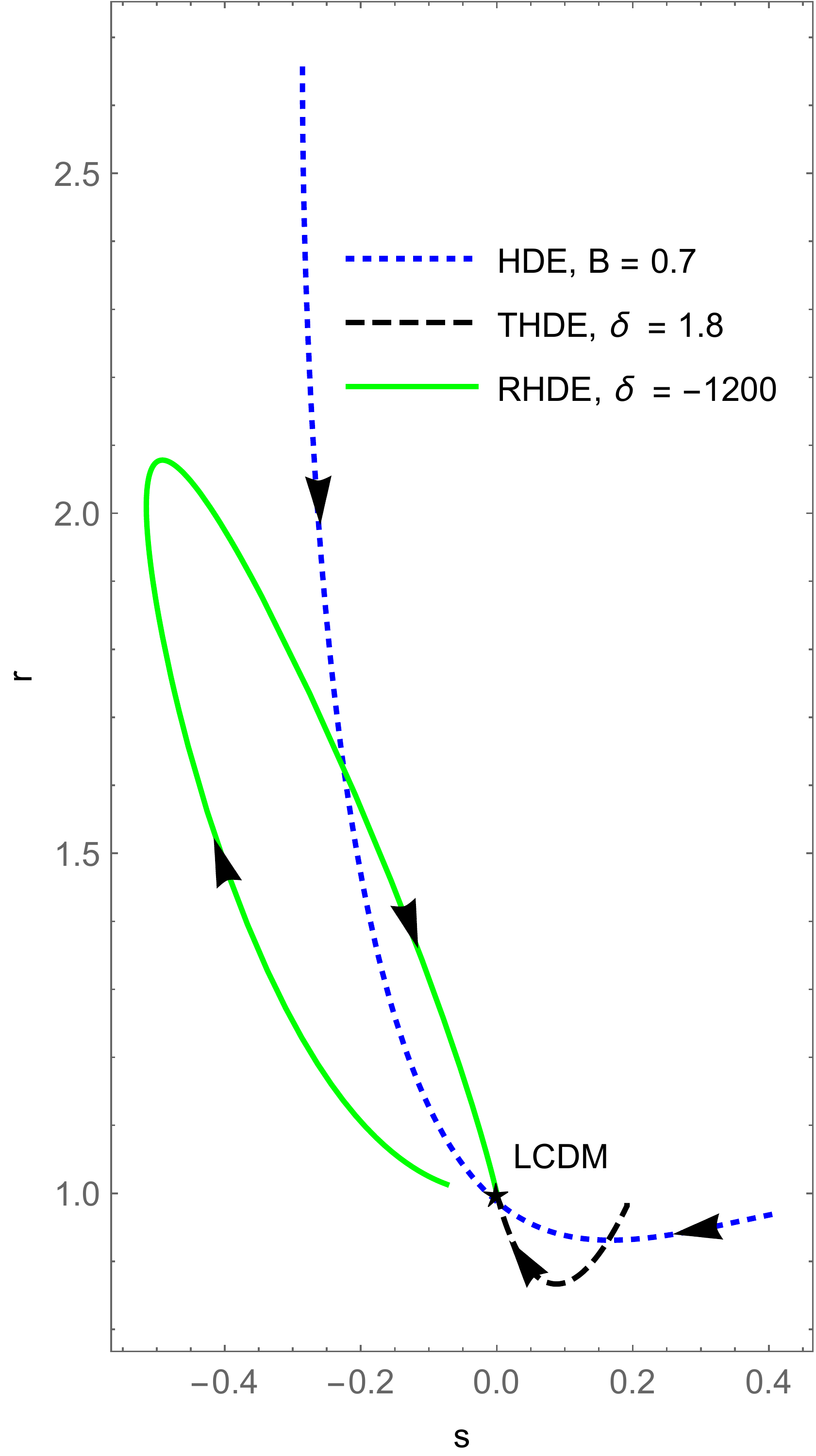}
			\caption{ Comparison of evolutionary trajectories  of the ($r, s$) pair plane for the HP inspired DE models.   Here, $\Omega_D(z=0)=0.73$, $H(z=0)= 70$, $\omega_{D}(z=0) = -
				0.90$.}
			\label{Omega-z8}
		\end{center}
	\end{figure}
	
	\subsection{The RHDE model} 
	Furthermore, one more  form of DE based on  R$\acute{e}$nyi entropy and  HP inspired DE has been  suggested  in \cite{ref33}, and the  energy density for the RHDE is given as
	
	\begin{eqnarray}
	\label{eq8}
	\rho _D=\frac{3 B^2 H^2}{8 \pi  \left(\frac{\pi  \delta }{H^2}+1\right)},
	\end{eqnarray}

 where the value of a numerical constant $ B^{2} $ is  taken as $B^2=\left(\frac{\pi  \delta }{H_0^2}+1\right) \left(1-\Omega _{{m0}}\right)$ \cite{ref30}. Taking the time derivative of $\Omega_D \equiv \frac{\rho_{D}}{ 3 M^{2}_{p}H^{2}}$ on both sides with respect to $x = log a $, we get
	
	\begin{equation}
	\label{eq9}
	\Omega _D'=-\frac{3 \pi B^2 \delta  H^2 \left(\Omega _D-1\right)}{\left(\pi  \delta +H^2\right) \left(\pi  \delta  \left(2 \Omega _D-1\right)+H^2 \left(\Omega _D-1\right)\right)},
	\end{equation}
	So, for the RHDE,  the EoS parameter is given as
	
	\begin{equation}
	\label{eq10}
	\omega _D=-\frac{\pi  \delta }{\pi  \delta  \left(2 \Omega _D-1\right)+H^2 \left(\Omega _D-1\right)},
	\end{equation}

	
	\begin{figure}[htp]
		\begin{center}
			\includegraphics[width=9cm,height=8cm]{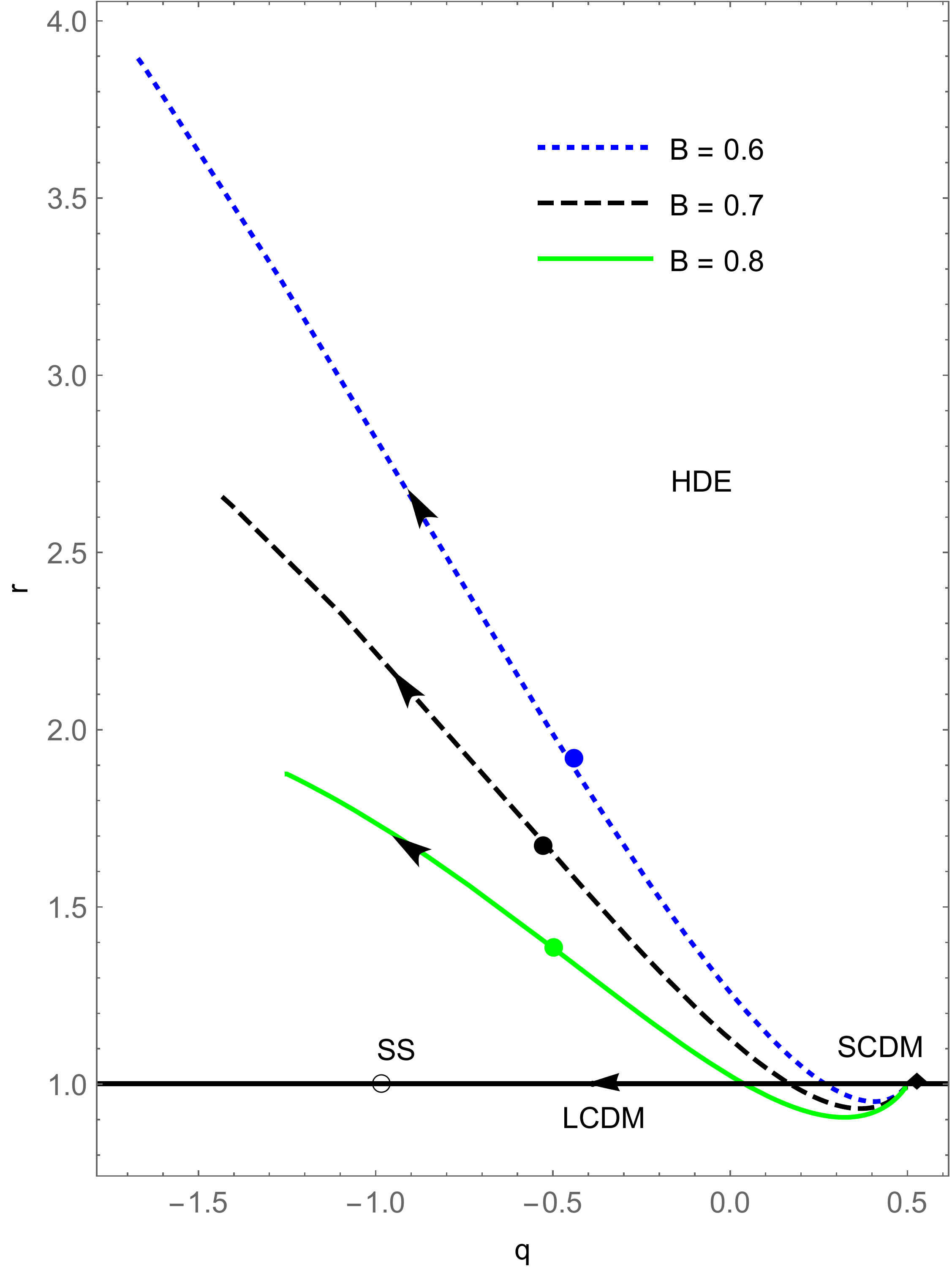}
			\includegraphics[width=9cm,height=8cm]{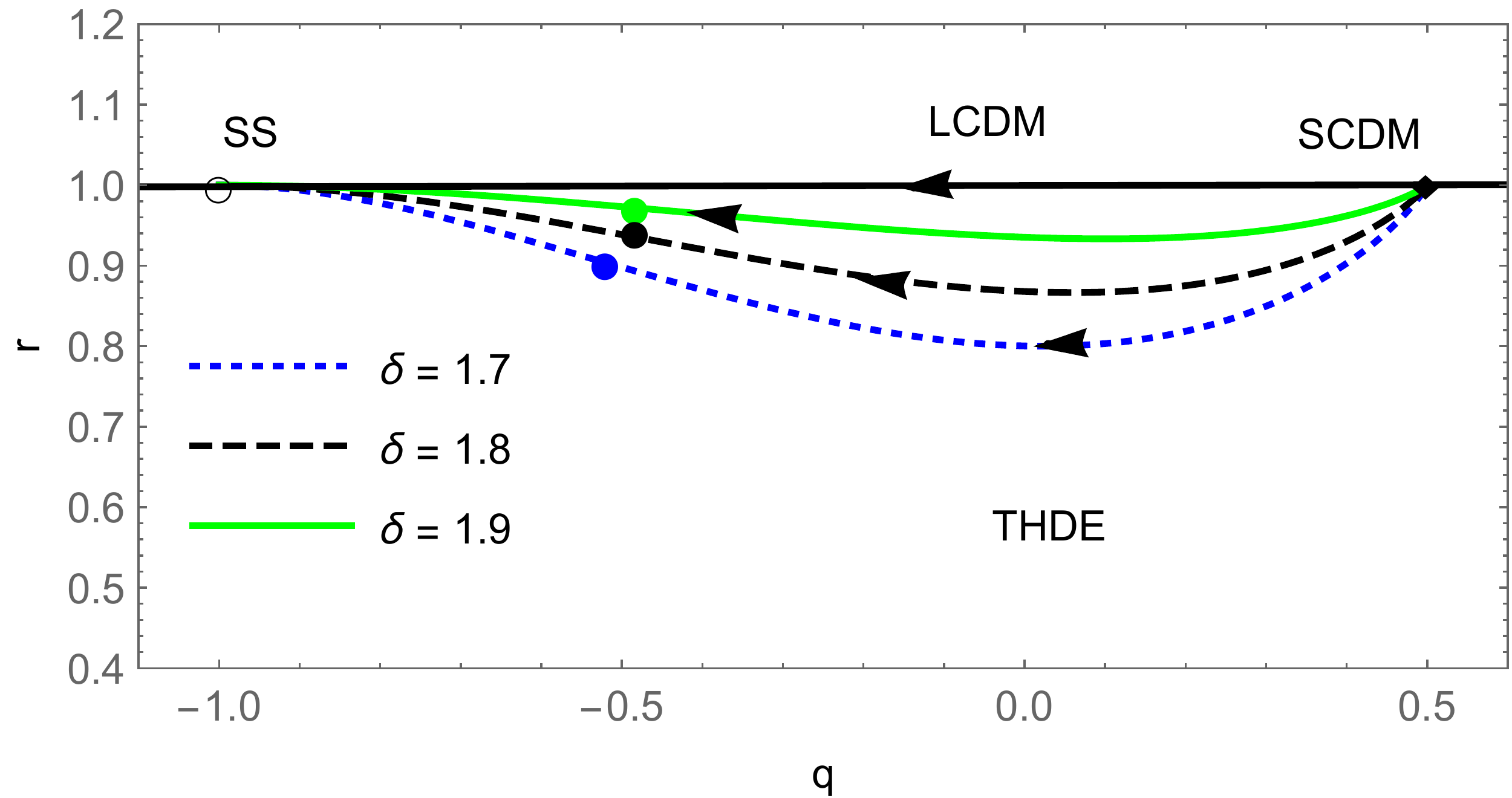}
			\includegraphics[width=9cm,height=8cm]{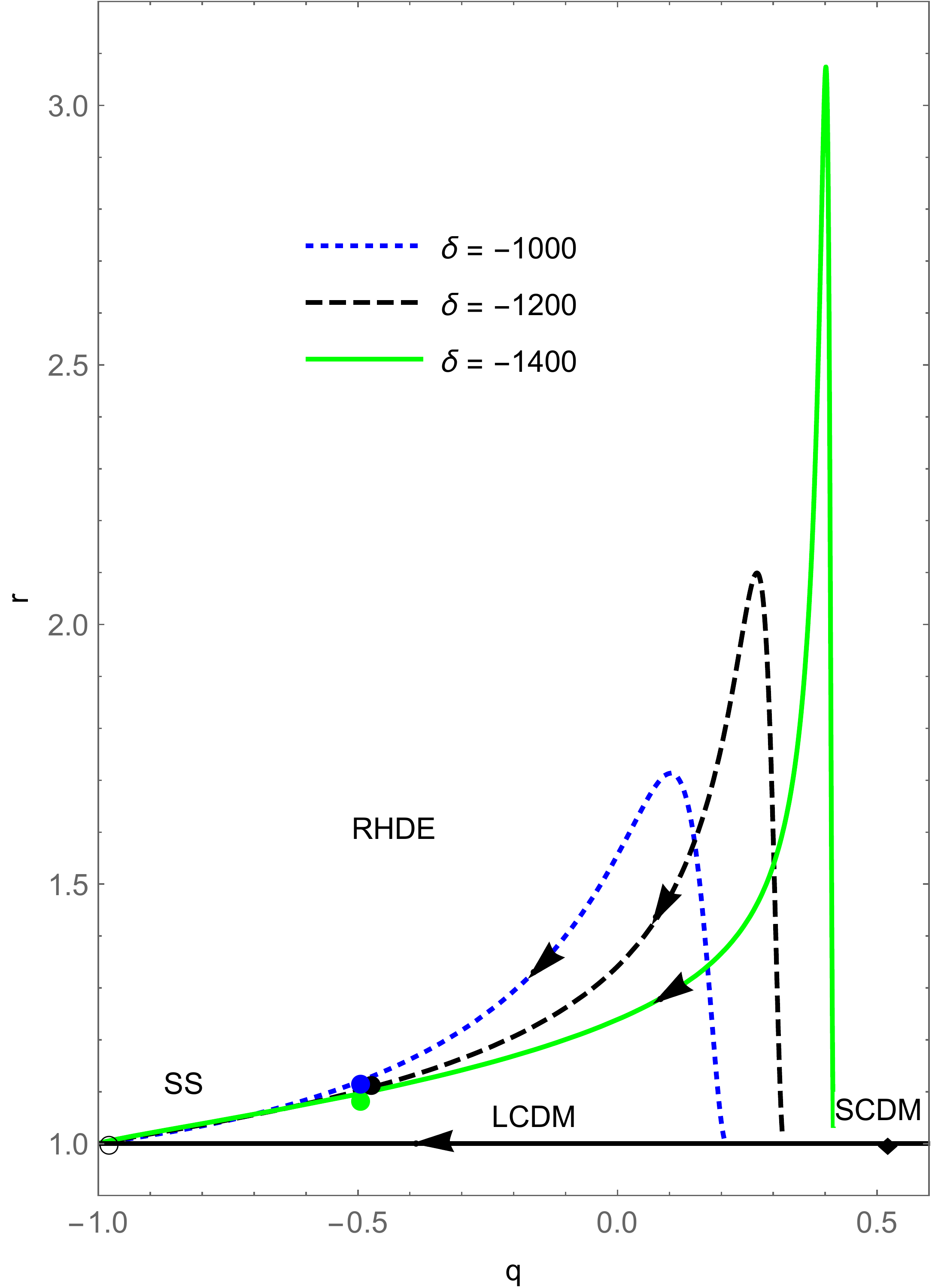}

			\caption{The evolutionary behaviour of the ($r, q$) pair for HDE, THDE and RHDE Model. Here, $\Omega_D(z=0)=0.73$, $H(z=0)= 70$, $\omega_{D}(z=0) = -
				0.90$.}
			\label{Omega-z10}
		\end{center}
	\end{figure}
	
	\begin{figure}[htp]
		\begin{center}
			\includegraphics[width=9cm,height=8cm]{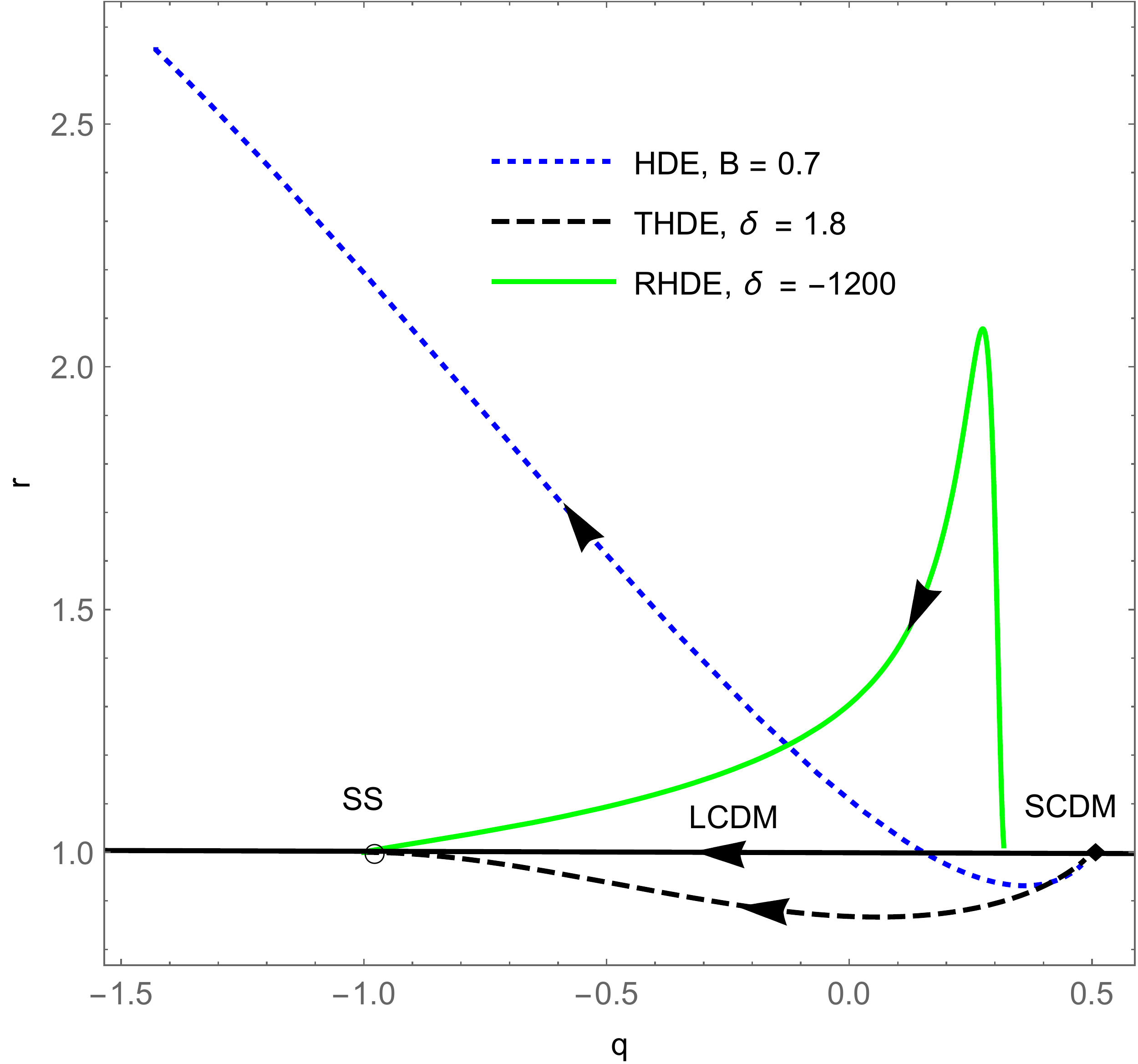}
			
			\caption{ Comparison of evolutionary trajectories  of the ($r, q$) pair for the HP inspired DE models.   Here, $\Omega_D(z=0)=0.73$, $H(z=0)= 70$, $\omega_{D}(z=0) = -
				0.90$.}
			\label{Omega-z10a}
		\end{center}
	\end{figure}

	\section{Statefinder diagnostic} 
	
	Before the presence of the statefinder parameters $(r)$, $(s)$ and the statefinder pair $(s, r)$ and $(q, r)$, 
	one may differentiate various  DE models with the help of
	the evolution of the deceleration parameter $q(z)$. Therefore,  we take into consideration the evolution of  $q$ for the aforementioned $3$ HP inspired DE models in this work.\\

 The deceleration
	parameter $q$ can be expressed in a spatially flat Universe in terms of $\Omega _D$ \cite{ref56}.

	\begin{equation}
	\label{eq11}
	q =  \frac{1}{2} + \frac{3 \omega _D \Omega _D}{2}
	\end{equation}
	
	 It is important to mention here that for all three models  i.e.  HDE, THDE and RHDE,	we fix $\Omega_D(z=0)=0.73$, $H(z=0)= 70$, $\omega_{D}(z=0) = -	0.90$. For the HDE model, parameter $B$ is taken as $(0.6, 0.7, 0.8)$ \cite{ref62,ref63}. For the THDE model, the model parameter $\delta$ is taken to be $(1.7, 1.8, 1.9)$ \cite{ref57,ref59}. The parameter $\delta$ in  the RHDE model is taken to be $(-1000, -1200, -1400)$ \cite{ref64}.\\

	The evolutionary behaviour of the deceleration parameter $q$ is plotted in Fig. 1, for all three HP inspired DE i.e. the HDE, THDE and RHDE models.
 We observe from the figure that for the HDE model the evolutionary behaviour of $q$ can be discriminated with each other for various values of the model parameter $B$ in the low-redshift region while it can not be discriminated in high- redshift region. For the THDE model, it is clear from the figure that the evolutionary behaviour of $q$ can not be discriminated with each other for different values of the model parameter $\delta$ in the low-redshift region as well as in high- redshift region. Also, for the RHDE model, it is clear from the figure that the evolutionary behaviour of $q$ can not be discriminated with each other for different values of the model parameter $\delta$ in the low-redshift region but can be differentiated  in high- redshift region. Although, the behaviour  of deceleration parameter $q$ for  all three DE model is in toe with the observational results, which depicting the deceleration to acceleration  phase of the universe. Similarly, we moreover graph the deceleration parameter $q$ evolution of the three HP inspired DE models in Fig. 2. 	From $q(z)$ diagnostic, it is difficult to differentiate among these DE models.\\

For discriminating among different DE models, now we investigate the statefinder parameter, which is constructed from the scale factor by taking its derivatives up to the third order. The statefinder pair $\left\lbrace r, s\right\rbrace$ is constructed from the space-time metric directly and it gives a very comprehensive overview of the	dynamics of the Universe and subsequently the nature of the DE.  The statefinder parameters are given as: \cite{ref52,ref53}
	\begin{eqnarray}
	\label{eq12}
	r = \frac{\dddot {a}}{aH^{3}}
	\end{eqnarray}
	\begin{eqnarray}
	\label{eq13}
	s = \frac{r-1}{3 (q - \frac{1}{2})}
	\end{eqnarray}
The statefinder parameters $r$ and $s$ can be expressed in terms of the EoS parameter and energy density as given below:
	\begin{eqnarray}
	\label{eq14}
	r= 1+ \frac{9}{2}\omega_{D}(1+\omega_{D})\Omega_{D}-\frac{3}{2}\omega_{D}^{'}\Omega_{D}
	\end{eqnarray}
	\begin{eqnarray}
	\label{eq15}
	s= 1+ \omega_{D}-\frac{1}{3}\frac{\omega_{D}^{'}}{\omega_{D}},
	\end{eqnarray}
	
	Where the prime represents the differential coefficient with  $x = \ln a $. The evolutionary behaviour of the first statefinder parameter $r$ with $z$ is shown in Fig. 3, for all three  HP inspired DE models with different model parameter values compared with $\Lambda$CDM model.  From Fig. 3 to Fig. 10, the arrows indicate the evolution directions.

  We see from Fig. 3, that for HDE model the evolutionary behaviour of the first statefinder parameter $r$ can be well discriminated from $\Lambda$CDM model ( denoted by black solid line $r = 1$ in the figure) in the region $ -1\leq  z\leq  1$  for different values of the model parameter $c$ while it can not be distinguished from  $\Lambda$CDM model in  the region $ 1\leq  z\leq 5$ . For the THDE and RHDE model, it is clear from the figure that the evolutionary behaviour of $r$  for different values of the model parameter $\delta$ is distinct from $\Lambda$CDM model in the region $ -0.5\leq  z\leq 5$ but can not be discriminated from $\Lambda$CDM model in the region $ -1\leq  z\leq -0.5$. A straight comparison  between the $\Lambda$CDM model and these three HP inspired DE models with the first statefinder parameter $r$ can be seen in Fig. 4. The differentiation of these three DE models are clearly observed in the region $ -0.5\leq  z\leq 5$. The difference between  the HDE model and the $\Lambda$CDM model becomes clearer in the range $ -1\leq  z\leq -0.5$, while THDE and RHDE models are not distinct from $\Lambda$CDM model in this range.\\

Fig. 5, shows the evolution of the second statefinder parameter
	$s$ versus $z$ of the three HP inspired DE models for different values of model parameters, also compared with the $\Lambda$CDM model  (denoted by black solid line $s = 0$ in the figure). For the HDE model, the evolutionary trajectories show legible separation  from the $\Lambda$CDM model in the region $ -1\leq  z\leq 5$. While, for the THDE and the RHDE models, the $s(z)$ curves are not distinct from $\Lambda$CDM model in the region  $ -1\leq  z\leq -0.5$, for different values of the model parameter $\delta$. These DE models separate more distinctly for different values of model parameters in the region $ -1\leq  z\leq -0.5$.  Fig. 6, compares the HDE, THDE and the RHDE models with the second statefinder parameter $s(z)$ from the $\Lambda$CDM model. The clear differentiation of these DE models can be seen in the region $ -0.5\leq  z\leq 5$. The evolutionary trajectories of both first and second statefinder parameters shows that these DE models are separate  distinctively in the region $ -0.5\leq  z\leq 5$, but THDE,  RHDE models are not differentiated and  approaches to  $\Lambda$CDM model in the region  $ -1\leq  z\leq -0.5$.\\
 

To break this degeneracy, the evolutionary trajectories for the HDE, THDE and the RHDE models are plotted in Fig. 7 for different parameter values in $(r, s)$  plane. The point $(s_0, r_0)$ denotes the present value of the statefinder parameters, which are marked as the dotted circle in the figures. The fixed point $(0, 1)$ presented by the star in these figures represents the $\Lambda$CDM.  We can discriminate DE models with the statefinder diagnostic $(r_0, s_0)$, if we have accurate information of $(r_0, s_0)$ which can be retrieved from the future high precision observational data.\\

		 We have quintessence if the parameters $(r<1, s>0)$, while if $(r>1, s<0)$, we have Chaplygin gas model.  From  Fig. 7 \& 8, we clearly observe that the trajectory of $(r, s)$  plane of HDE model shows  quintessence as well as Chaplygin gas behaviour  at an early time for different values of model parameter  $c$ and finally reaches to  $\Lambda$CDM  in the far future. The evolution of the  $(r, s)$  plane for the THDE  model shows that the curve lies in the quintessence region at early time and finally approaches to $\Lambda$CDM at late time. While the evolutionary trajectories of the statefinder pair $(r-s)$ of the RHDE model show only the Chaplygin gas behaviour at an early time and forms swirl before reaching to $\Lambda$CDM in the remote future. Which is quite different from the other two DE models.  Fig. 7, also depicts that the model parameters of these DE models not only discriminate the present values of $\left\lbrace s, r\right\rbrace $ but also differentiate the evolutionary trajectory in $(r, s)$  plane.\\

	 As a complementarity for the diagnostic, we also plot the evolutionary trajectories of the statefinder pair ($q-r$) in Fig. 9. The point $(0.5, 1)$ represents the SCDM, that is the  matter-dominated Universe in this graph, and the fixed point $(-1, 1)$ represents 
	SS- the de Sitter expansion i.e. the steady-state, which are marked by filled diamonds and the empty circle, respectively.  Important point to mention here that the black solid horizontal line 
	starting from the fixed point of SCDM model  and ending at the fixed point of SS model denotes evolution of the $\Lambda$CDM model, which divides the $q-r$  plane into $2$ parts. The upper half is occupied by Chaplygin 
	gas model and the lower half contain quintessence models.  We also observe the signature flip in the value of q from +ve to -ve, which explains
		the recent phase transition successfully. For the HDE model, we can see that both the LCDM scenario and HDE model start evolving from the same point in the past $(r = 1, q = 0.5)$
	which corresponds to a matter-dominated SCDM Universe, and crosses the point $(q = -1, r = 1)$ in the future, which corresponds to a steady-state cosmology (SS)—the de Sitter expansion.  For the  THDE model evolutionary trajectories start evolving from  SCDM i.e. matter-dominated Universe ($q = 0.5$, $r = 1$) in the past, and  their evolution ends at the same point ($q = -1$, $r = 1$) in the future for 	different parameter values.  For the  RHDE model, evolutionary trajectories start evolving from different points in the past, and  their evolution ends at the point ($q = -1$, $r = 1$) in the future for different parameter values. Thus, the `distance' from this model to the de Sitter 
	expansion ($SS$) can be easily identified in this diagram.  A direct comparison of these DE	models in the ($q-r$)  plane is shown in Fig. 10. 
	
 The evolution trajectories of the statefinder pair $(r-q)$ of the HDE model show both quintessence and the Chaplygin gas behaviour.  The THDE model always lies in the quintessence region.  The evolutionary trajectory of the RHDE model corresponds to the Chaplygin gas model.\\

	\section{Concluding remarks}
	
	Many dynamical DE models have been proposed so far in the literature, inspired by the HP and various entropy formalisms. Hence, it is of great interest to differentiate these HP inspired DE models with
	the observational results. Although, especially, in the	low-redshift region, these models are
	degenerate with each other to some extent. We compare three important
	HP inspired DE models, i.e., the THDE, HDE and RHDE and we use the statefinder parameter analysis to differentiate
	them  in this work.\\
	
	We  observe from the $q(z)$ evolution that THDE and RHDE models can not be discriminated in the low-redshift region while HDE model can be discriminated in this region. Interestingly, all three DE models represent the decelerating to accelerating universe for different parameter values. Since most of the observational results are mainly within the low red-shift region (generally $z\leq1$).  For the more clear discrimination among these DE models in the low-redshift region, it is important to use some quantities with higher-order differential coefficients of the scale factor.  We observe that the	first and second statefinder parameters $r (z)$ and $s(z)$ are very helpful for this purpose.

	We  observe from the $r(z)$ and  $s(z)$ analysis that THDE and RHDE models can not be differentiated in the low-redshift region from the  $\Lambda$CDM model but can be discriminated in the range $-0.5<z<5$ from the  $\Lambda$CDM model. It is clear that both the THDE and RHDE approach to the $\Lambda$CDM model in the low-redshift region.  The RHDE and HDE models can not be discriminated in the high-redshift region from the $\Lambda$CDM model in $r(z)$ plane, while all the three DE models can be well discriminated from  $\Lambda$CDM model in $s(z)$ plane for the various parameter values in the high red-shift region. \\
	
	For all three HP inspired DE models i.e. THDE, HDE and RHDE a direct comparison in	$(r, s)$ and  $(r, q)$ plane has also been done, in which the discrimination between these three models with the $\Lambda$CDM model may be easily seen.  The HDE model shows both quintessence and Chaplygin gas behaviour. The THDE model always lies in the quintessence region while the RHDE model shows the Chaplygin gas behaviour. We hope that in future the model parameters of each DE model can be obtained upon a comparison against several low-redshift observational data  such as  Pantheon SN Ia data, baryon acoustic oscillations (BAO) and cosmic chronometers (CC)
		  \cite{ref65,ref66,ref67,ref67a}.\\

	In this work, only statfinder diagnostic tool is used to discriminate the HDE, THDE and RHDE models. Some other diagnostic tools are also available in the literature such as : statefinder hierarchy, growth rate of perturbations, Om diagnostic and $\omega-\omega^{\prime}$ pair, swampland conjectures \cite{ref68}.

	 In future work, it is fascinating to make use of these diagnostic tools to differentiate and  shed light on the behaviour of various DE models.


	\section*{Acknowledgements}
	The authors are highly indebted to the GLA University, India for the support in this academic endeavour.


\end{document}